\documentstyle[psfig]{mn}

\newif\ifAMStwofonts
\ifoldfss
  \newcommand{\rmn}[1] {{\rm #1}}

  \ifCUPmtlplainloaded \else
    \NewTextAlphabet{textbfit} {cmbxti10} {}
    \NewTextAlphabet{textbfss} {cmssbx10} {}
    \NewMathAlphabet{mathbfit} {cmbxti10} {} % for math mode
    \NewMathAlphabet{mathbfss} {cmssbx10} {} %  "   "    "
  \fi
  \ifAMStwofonts
    \ifCUPmtlplainloaded \else
      \NewSymbolFont{upmath} {eurm10}
      \NewSymbolFont{AMSa} {msam10}
      \NewMathSymbol{\upi}     {0}{upmath}{19}
      \NewMathSymbol{\umu}     {0}{upmath}{16}
      \NewMathSymbol{\upartial}{0}{upmath}{40}
      \NewMathSymbol{\leqslant}{3}{AMSa}{36}
      \NewMathSymbol{\geqslant}{3}{AMSa}{3E}

    \fi
  \fi
\fi % End of OFSS

\ifnfssone
  \newmathalphabet{\mathit}
  \addtoversion{normal}{\mathit}{cmr}{m}{it}
  \addtoversion{bold}{\mathit}{cmr}{bx}{it}
  \newcommand{\rmn}[1] {\mathrm{#1}}

  \newmathalphabet{\mathbfit} % math mode version of \textbfit{..}
  \addtoversion{normal}{\mathbfit}{cmr}{bx}{it}
  \addtoversion{bold}{\mathbfit}{cmr}{bx}{it}
  \newmathalphabet{\mathbfss} % math mode version of \textbfss{..}
  \addtoversion{normal}{\mathbfss}{cmss}{bx}{n}
  \addtoversion{bold}{\mathbfss}{cmss}{bx}{n}
  \ifAMStwofonts
    \ifCUPmtlplainloaded \else
      \UseAMStwoboldmath
      \makeatletter
      \new@mathgroup\upmath@group
      \define@mathgroup\mv@normal\upmath@group{eur}{m}{n}
      \define@mathgroup\mv@bold\upmath@group{eur}{b}{n}
      \edef\UPM{\hexnumber\upmath@group}
      \new@mathgroup\amsa@group
      \define@mathgroup\mv@normal\amsa@group{msa}{m}{n}
      \define@mathgroup\mv@bold\amsa@group{msa}{m}{n}
      \edef\AMSa{\hexnumber\amsa@group}
      \makeatother
      \mathchardef\upi="0\UPM19
      \mathchardef\umu="0\UPM16
      \mathchardef\upartial="0\UPM40
      \mathchardef\leqslant="3\AMSa36
      \mathchardef\geqslant="3\AMSa3E
    \fi
  \fi
\fi % End of NFSS release 1

\ifnfsstwo
  \newcommand{\rmn}[1] {\mathrm{#1}}

  \DeclareMathAlphabet{\mathbfit}{OT1}{cmr}{bx}{it}
  \SetMathAlphabet\mathbfit{bold}{OT1}{cmr}{bx}{it}
  \DeclareMathAlphabet{\mathbfss}{OT1}{cmss}{bx}{n}
  \SetMathAlphabet\mathbfss{bold}{OT1}{cmss}{bx}{n}
  \ifAMStwofonts
    \ifCUPmtlplainloaded \else
      \DeclareSymbolFont{UPM}{U}{eur}{m}{n}
      \SetSymbolFont{UPM}{bold}{U}{eur}{b}{n}
      \DeclareSymbolFont{AMSa}{U}{msa}{m}{n}
      \DeclareMathSymbol{\upi}{0}{UPM}{"19}
      \DeclareMathSymbol{\umu}{0}{UPM}{"16}
      \DeclareMathSymbol{\upartial}{0}{UPM}{"40}
      \DeclareMathSymbol{\leqslant}{3}{AMSa}{"36}
      \DeclareMathSymbol{\geqslant}{3}{AMSa}{"3E}
    \fi
  \fi
\fi % End of NFSS release 2

\ifCUPmtlplainloaded \else
  \ifAMStwofonts \else % If no AMS fonts
    \def\upi{\pi}
    \def\umu{\mu}
    \def\upartial{\partial}
  \fi
\fi

\def\hi {\hbox{H\,{\sc i}}}

\def\kmss{km~s$^{-1}$ }
\def\kms{km~s$^{-1}$}

\def\disp{\ifmmode {{\langle}v^2{\rangle}^{1/2}}
           \else {${\langle}v^2{\rangle}^{1/2}$} \fi}
\def\dispr{\ifmmode {{\langle}v^2_R{\rangle}^{1/2}}
           \else {${\langle}v^2_R{\rangle}^{1/2}$} \fi}
\def\disprn{\ifmmode {{\langle}v^2_R{\rangle}^{1/2}_{R=0}}
           \else {${\langle}v^2_R{\rangle}^{1/2}_{R=0}$} \fi}
\def\disprh{\ifmmode {{\langle}v^2_R{\rangle}^{1/2}_{R=h}}
           \else {${\langle}v^2_R{\rangle}^{1/2}_{R=h}$} \fi}
\def\dispt{\ifmmode {{\langle}v^2_{\Theta}{\rangle}^{1/2}}
           \else {${\langle}v^2_{\Theta}{\rangle}^{1/2}$} \fi}
\def\dispz{\ifmmode {{\langle}v^2_z{\rangle}^{1/2}}
           \else {${\langle}v^2_z{\rangle}^{1/2}$} \fi}
\def\dispzn{\ifmmode {{\langle}v^2_z{\rangle}^{1/2}_{R=0}}
           \else {${\langle}v^2_z{\rangle}^{1/2}_{R=0}$} \fi}
\def\vmaxd{\ifmmode {v^{\rmn disc}_{\rmn max}}
           \else {$v^{\rmn disc}_{\rmn max}$} \fi}
\def\vmaxds{\ifmmode {v^{\rmn disc}_{\rmn max}}
           \else {$v^{\rmn disc}_{\rmn max}$ } \fi}
\title[Structure and kinematics of NGC 6503]
{An investigation of the structure and kinematics of the spiral
galaxy NGC 6503}

\author[Roelof Bottema and Jeroen P.E. Gerritsen]
       {Roelof Bottema and Jeroen P.E. Gerritsen \\
        Kapteyn Astronomical Institute, P.O. Box 800, 
        NL-9700 AV Groningen, the Netherlands}
\date{Accepted date 1.
      Received date 2;
      in original form date 3}

\pagerange{\pageref{firstpage}--\pageref{lastpage}}
\pubyear{1997}

\begin{document}

\maketitle

\label{firstpage}

\begin{abstract}
The spiral galaxy NGC 6503 exhibits a regular kinematical structure
except for a remarkable drop of the stellar velocity
dispersion values in the central region.
To investigate the dynamics of the disc in general, and
that of the central region in particular, a theoretical framework
has been described. This includes a mass decomposition of the
galaxy into a family of disc/halo realizations compatible
with the observed photometry and rotation curve. For this family
stellar velocity dispersion values and stability parameters
were calculated, showing that the more massive discs,
although having larger dispersions, are less stable.
However, a reliable theoretical description of the inner regions
where the drop occurs cannot be given.

That is why we have resorted to numerical calculations.
Not only to study the central region, but also to investigate
the appearance of the disc in a general sense. Pure stellar
3d simulations have been performed for the
family of decompositions. A clear result is that
disc/dark halo mass ratios approaching
those of the maximum disc limit generate
a large bar structure. This is incompatible with the
observed morphology of NGC 6503.
For radii larger than $\sim$ 0.2 scalelengths, the stellar
kinematics resulting from the simulations essentially
agrees with that predicted by the theory.
But, unfortunately, the central velocity
dispersion drop could not be reproduced.

A close inspection reveals that the central nuclear region is
very small and bright. Therefore, tentatively, this nucleus was
considered as an isothermal sphere and a core fitting
procedure was applied.
For an adopted equal mass-to-light ratio of disc and nucleus,
a velocity dispersion of 21.5 \kmss is predicted, in excellent agreement
with the observed central value.
An analysis, in retrospect, of the local densities involved proves that
the nucleus is local and gravitationally dominant such that its
approximation as an isothermal sphere, is justified. The
observed dispersion drop can thus be explained by a separate
kinematically distinct galactic component.
\end{abstract}

\begin{keywords}
galaxies: individual: NGC 6503 -- galaxies: haloes -- galaxies: kinematics
and dynamics -- galaxies: spiral -- galaxies: structure -- 
methods: numerical.
\end{keywords}

\section{Introduction}

Stellar velocity dispersions provide a direct measure of the local
densities in a galactic disc. These densities again determine the
local and global stability against spiral arm and bar disturbances
in a disc (Toomre 1964; Athanassoula et al. 1987). 
Unfortunately discs are faint
and some effort has to be spent extracting the desired dispersion 
values. For a number of discs this has now been done (van der Kruit \&
Freeman 1984, 1986; Lewis \& Freeman 1989; Bottema 1995, and references
therein). For one galaxy, NGC 6503 (Bottema 1989, hereafter B89)
the radial dispersion functionality did not follow the expected 
proportionality with the square root of the disc surface density. 
Instead in the central region, for $R \la$  0.3 radial scalelengths,
there was a distinct drop of the dispersion values. 
This drop is well established because the galaxy is bright
at those positions, resulting in small errors
of the dispersion, and lineprofiles were sufficiently
resolved by the employed spectrographic setup.
Any effects of dust are always small because
the stellar dispersions are measured
of an old stellar population having a local
isothermal distribution. 
This study set out to explain this central dispersion drop.

NGC 6503 is a regular inclined spiral galaxy of moderate size.
It shows modest spiral arm features but no grand design structures.
Neutral hydrogen observations were performed by a.o. Wevers et al.
(1986) and Begeman (1987), who derived a regular rotation curve
reaching 120 \kmss and remaining at a constant level of 116 \kmss out to large
radii. Surface photometry by Wevers et al. (1986) and B89 showed that
the galaxy can be described by an exponential radial profile with
a scalelength $(h)$ of 40 arcsec for $R <$ 160 arcsec and $h$ = 80 arcsec
for $R >$ 160 arcsec with a nuclear region at the centre confined to 
$R \sim$ 8 arcsec. Relevant parameters are summarized in Table 1
and for convenience a photograph of the galaxy is presented in Fig. 1.

\begin{figure}
%\vspace{10cm}
\psfig{figure=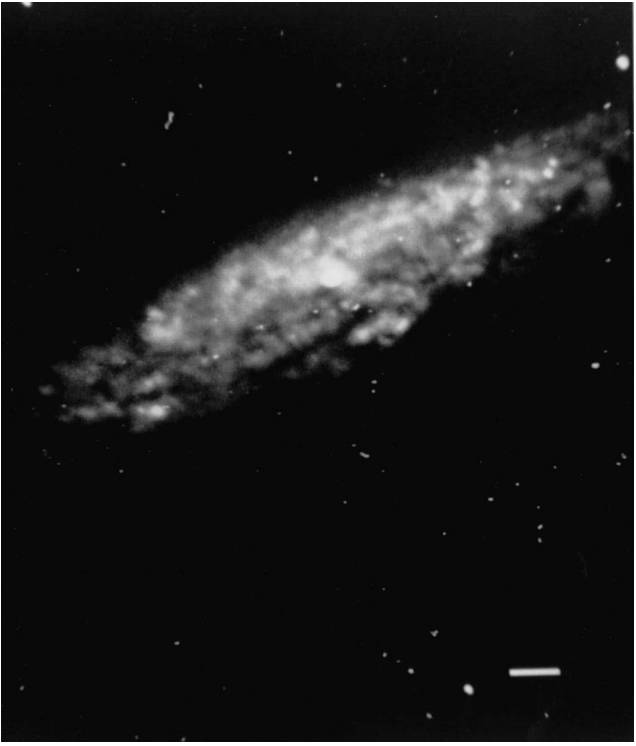,width=8.4cm}
\caption{
Optical image of NGC 6503 in the B-band copied from Bottema (1989).
The size of the image on the sky is 134 arcsec $\times$ 215 arcsec; the line
in the lower right corner has a length of 10 arcsec. North at top,
East on left.}
\end{figure}
Bottema (1993, hereafter B93) suggested an explanation of
the observed drop. It involves the formalism devised by Bahcall 
(1984) to describe a stellar disc consisting of multiple stellar components 
embedded in a dark halo. The full global application of the relevant 
equations allows, for certain conditions which might apply for
NGC 6503, that a significant cooling of the stars may occur in the 
central regions. In Sect. 2 this procedure is described.
The mass distribution of the galaxy is determined by the observed
rotation curve and photometry, except for the ratio of disc to dark halo
masses (van Albada et al. 1985). Usually astronomers chose
a certain fixed ratio according to their personal preferences.
Only a few studies exist which investigate the effect on the galaxy
of changing this ratio (Athanassoula et al. 1987). To cover all possibilities
in Sect. 3 for NGC 6503 a family of disc/halo ratios is constructed.
For this family velocity dispersion values and stability parameters
are calculated.

Theoretical descriptions of the stellar kinematics of galactic discs
have been given in various ways. Examples are the approach by Bahcall
just mentioned, an analysis of higher moments of the Boltzmann 
equation (Amendt \& Cuddeford 1991) or by starting out with a suitable
choice of the distribution function (Kuijken \& Dubinski 1995). All these
descriptions use certain assumptions and approximations being more or less
valid. To avoid such approximations,
in Sect. 4 numerical calculations have been performed
for the family of disc/dark halo mass ratios constructed for NGC 6503.
Simulated stellar velocity dispersions are compared with the observations
and with the theoretically predicted values. In addition, the morphology 
of the disc is investigated as a function of the disc/halo mass ratio.

Although the theoretical and numerical calculations provide considerable
insight into various aspects of the disc kinematics of NGC 6503, a
consistent explanation of the central dispersion drop could not be found.
In Sect. 5, first tentatively and later confirmed, the explanation
is given as the result of a separate, kinematically distinct 
galactic component.

\setcounter{table}{0}
\begin{table}
\caption{Parameters of NGC 6503}
\begin{tabular}{lll}
\hline
 R.A. (1950) & 17$^{\rmn h}$ 49$^{\rmn m}$ 58$^{\rmn s}$ & $^{\rmn a}$ \\
 Declination (1950) & 70\degr {09}\farcm{5} &  $^{\rmn a}$ \\
 Hubble type & Sc(s) II.8 & $^{\rmn a}$ \\
 Inclination & 74\degr & $^{\rmn b}$ \\
 Pos. angle major axis & 121\degr & $^{\rmn b}$ \\
 Max. rotational vel. & 120 \kms & $^{\rmn b}$ \\
 \hi\/ systemic vel. & 26 \kms & $^{\rmn b}$ \\
 Total \hi\/ mass & 1.6 10$^9$ $M_{\sun}$ & $^{\rmn b}$ \\
 Phot. scale length & 40 arcsec for $R <$ 160 arcsec & $^{\rmn c}$ \\
    &                 80 arcsec for $R >$ 160 arcsec & \\
 Distance & 6 Mpc & $^{\rmn d}$ \\
 Scale & 1 kpc = {34}\farcs{4} &  \\
\hline
 \multicolumn{3}{l}{\quad $^{\rmn a}$ Sandage \& Tammann (1981)} \\
 \multicolumn{3}{l}{\quad $^{\rmn b}$ Begeman (1987)} \\
 \multicolumn{3}{l}{\quad $^{\rmn c}$ Wevers et al. (1986)} \\
 \multicolumn{3}{l}{\quad $^{\rmn d}$ Adopted} \\
\end{tabular}
\end{table}

\section[]{The velocity dispersion of a\\* stellar disc}

\subsection{The z-dispersion of an isolated disc}

For a locally isothermal, isolated stellar disc the vertical
density functionality $\rho(R,z)$ is found to be (Spitzer 1942)

\begin{equation}
  \rho(R,z) = \rho(R,0)\; {\rmn sech}^2\left( {{z}\over{z_0}}\right),
\end{equation}
with a defined vertical stellar velocity dispersion
${\langle}v^2_z{\rangle}^{1/2}_{\rmn isol}$ given by

\begin{equation}
  {\langle}v^2_z{\rangle}^{1/2}_{\rmn isol} = 
  \sqrt{ \pi G \sigma(R) z_0 }.
\end{equation} 
In this equation $\sigma(R)$ is the surface density of the disc and
$z_0$ a scaleheight which is found to be practically constant as a
function of radius (van der Kruit \& Searle 1981a, b, 1982; de Grijs
\& van der Kruit 1996).

\subsection{Embedding in a dark halo}

A real stellar disc is not isolated, but is expected to be embedded in a
massive dark halo. Furthermore the local approximation does not hold
everywhere; forces originating at other parts of the disc act upon the
position at $R,z$. This general situation has been studied by Bahcall (1984)
and applied by him to the solar neighbourhood. For a whole disc
consisting only of one stellar component embedded in a dark 
halo it can be shown (B93) that the relation between the vertical velocity
dispersion \dispz, surface density and scaleheight $z_0$ becomes

\begin{equation}
  \dispz = f(\varepsilon(R))\cdot \sqrt{ \pi G \sigma(R) z_0},
\end{equation}
Here $\varepsilon$ is a dimensionless parameter
given by

\begin{equation}
  \varepsilon(R) = {{ {\rho}_{\rmn H}^{z=0} - {{1}\over{4\pi G R}}
  {{\partial}\over{\partial R}} v^2_{\rmn c} }\over
  {{\rho}_{\rmn D}^{z=0} }},
\end{equation}
expressing the relative importance of disc, halo, and
non locality. The total rotational velocity is $v_c$
and the space densities in the plane of the galaxy of the halo
and the disc are
${\rho}_{\rmn H}^{z=0}$ and ${\rho}_{\rmn D}^{z=0}$ 
respectively.
For large ratios of disc to halo masses $\varepsilon$ becomes larger 
than zero and $f(\varepsilon)$ larger than one. This results in larger
dispersion values of a disc embedded in a dark halo compared to
the dispersion of an isolated disc. Near the central region of a galaxy
a rising rotation curve can produce negative $\varepsilon$ values
(Eq. 4) giving rise to $f(\varepsilon)$ values less than one.
So, near the centre of a galaxy the dispersions can become smaller
than those expected for an isolated locally isothermal disc.

In B93 a calculation is given of $f(\varepsilon)$. Unfortunately that
calculation is incomplete, leading qualitatively to the right results
(as described above), but quantitatively to wrong values of $f(\varepsilon)$.
In a recent study (Bottema, in preparation) a more complete analysis
is made of the situation when a disc is, or is being embedded in a halo.
That analysis is quite complicated and beyond the scope of this paper.
Still, one of the results is relevant for matters investigated in Sect. 4.
When an isolated disc is embedded in a dark halo the
stars will resettle and the vertical density profile
will change. That happens in such a way that the vertical velocity
dispersion slightly increases and the disc becomes thinner. The increase
in dispersion is 15 per cent for an initial (before resettling) value of 
$\varepsilon = 1$. Hence, for realistic disc/halo mass ratios there is
just a small increase of the dispersion only in the outer regions, when
a disc is allowed to evolve from locally isothermal to the true
equilibrium situation. 

\subsection[]{How to explain the central velocity dispersion\\* drop}

The observed stellar velocity dispersions of NGC 6503 show a
remarkable drop in the central regions (Fig. 2). It was argued
in B93 that this drop might be explained by the central non-locality
leading to appreciable negative $\varepsilon$ values and consequently
to a  dispersion much smaller than that of an isolated disc.
This has been investigated but a problem turned up.
The region where the drop occurs is typically
for radii smaller then the $z_0$ scaleheight. Just at those positions
a number of assumptions going into the calculations break down.
For example the plane parallelity of the disc and constantness
of the halo density (see Bahcall, 1984, for an extensive description)
are not valid any more. This leads to the conclusion that the
$\varepsilon$ calculations cannot be used to explain the central
dispersion drop.

\begin{figure}
%\vspace{8.5cm}
\psfig{figure=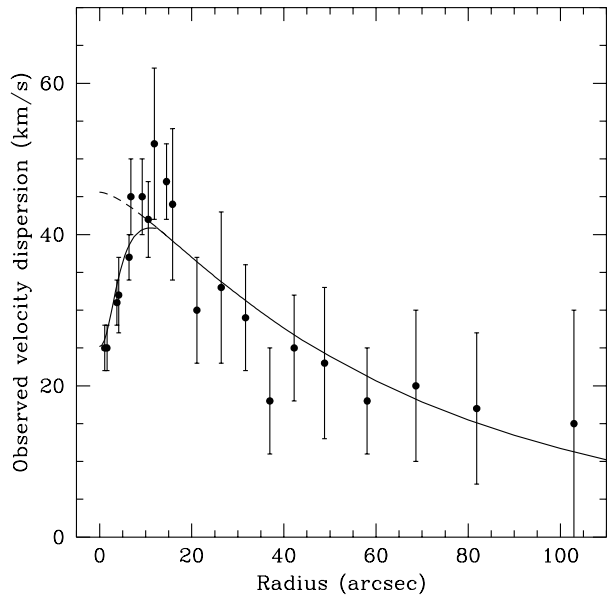,width=8.4cm}
\caption{The observed stellar velocity dispersions along the major axis
of NGC 6503. Note the remarkable drop near the centre of the galaxy.
The line gives the predicted dispersion for a disc plus nucleus
as described in Sect. 5, while the dashed line shows the
dispersion expected for a disc only.
}
\end{figure}
At the central regions of NGC 6503 the observable velocity dispersion
is dominated by the dispersion in the tangential and radial directions.
If the scaleheight and mass-to-light ratio do not change
drastically going inwards, at least the vertical dispersion
cannot decrease sharply near the centre of the galaxy.
Then the drop should be explained by a strongly changing ratio
of vertical to tangential (and radial) dispersion.
Such a scenario is not very likely because stellar heating
mechanisms show a preference for nearly equal ratios of the dispersions
(Binney \& Lacey 1988). Another way to explain the central drop is by
a significantly lower $z_0$ value at small radii. Such a very thin disc
near the centre of a galaxy might be possible and would go unnoticed
in photometric observations of edge-on galaxies. But, in order
to achieve this in a galactic disc, stellar heating must have been 
absent, which is not very likely

Anyway, a consistent analytical description
of the very central regions does not exist or is fraught with other
doubtful assumptions. This lead us to employ numerical simulations
of the NGC 6503 situation, not only to see if the drop could be
reconstructed, but also to
investigate the disc stability for different
disc to dark halo mass ratios.

\section{A family of disc-halo realizations}

\subsection{The principle}

For spiral galaxies there exists a fundamental uncertainty: the 
relative contribution of the mass of the luminous disc to the
total mass is not known from basic principles, nor can it be
determined from a rotation curve analysis (van Albada et al. 1985).
Stellar velocity dispersion observations of a sample of
galactics discs (B93) indicate that a disc has a mass such that the
maximum rotation of the disc is, on average, 63 per cent of the
observed maximum rotation. On the other hand, arguments have been presented
favouring the ``maximum disc hypothesis'' (van Albada \& Sancisi 1986;
Freeman 1992) which states that the disc provides a maximum
possible contribution to the observed rotation.
Therefore, for the numerical calculations which follow, a number
of disc/dark halo mass ratios is considered, covering the range
between the 63 per cent criterion and the maximum disc situation.
For this family of disc-halo realizations the galaxy kinematics
and disc morphology will be investigated.

\begin{figure}
%\vspace{10.7cm}
\psfig{figure=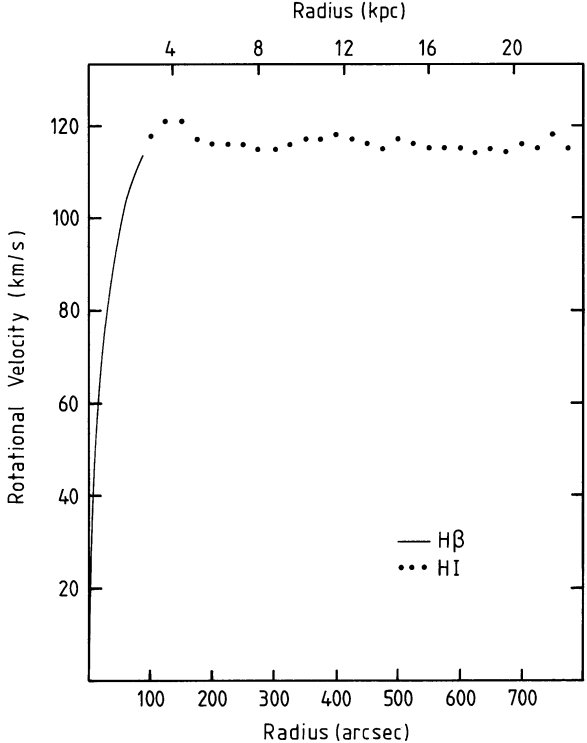,width=8.4cm}
\caption{Rotation curve of NGC 6503. A smooth curve has been fitted to the
H$\beta$ data (B89) for the inner regions. \hi\/ data points are
from Begeman (1987).
}
\end{figure}
As fundamental input the observed rotation curve (Fig. 3) is taken.
For the mass distribution we adopt a radially exponential sech$^2$
disc, following from a fit to the observed R-band photometry (Fig. 4),
thus assuming a constant mass-to-light ratio. This exponential
disc was assigned four different M/L ratios, resulting in four disc
masses and four different values of the maximum disc rotation
(\vmaxd) of 68, 80, 90, and 100 \kms. The disc rotation curve was 
calculated (Casertano 1983) and subtracted from the observed
rotation. The rotation which remains was considered to be generated
by the dark halo. A fit was made to this remaining rotation
for a suitable halo density distribution resulting in four different
dark halo situations. Such a scheme will also be used in a future
paper where stability and appearance of spiral structure of a more
realistic galaxy will be studied.

\begin{figure}
%\vspace{6.5cm}
\psfig{figure=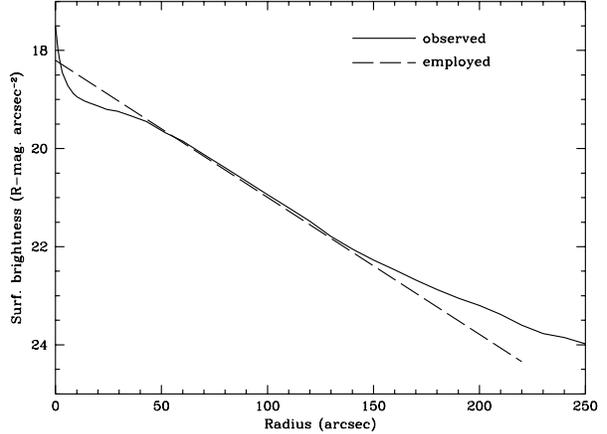,width=8.4cm}
\caption{The observed radial light profile in the R-band (B89), given
by the full drawn line. The dashed line indicates the exponential
disc actually used in the analyses. Between radii of 5 to 40 arcsec
this approximation may be considered as an absorption correction, which can be
justified by the redder colour at those positions.
}
\end{figure}

\subsection{Approximation by an exponential light profile}

As can be seen in Fig. 4, there is a plateau in the photometry between
radii of 5 to 40 arcsec. This plateau has been replaced by an exponential
profile all the way in, which is, of course, convenient for the
calculations but may at first glance seem ad-hoc. It can be
justified by three reasons. First between 5 to 40 arcsec the disc
is redder than further out (B89) suggesting the presence of dust which
obscures a part of the light. The replacement by an exponential
now serves as an absorption correction. Secondly, the photometry
as observed can be used to construct a disc rotation curve. When
this disc rotation is subtracted from the total rotation, a halo rotation
remains which is unrealistic and even unphysical for the more
massive disc situations, as illustrated in Fig. 5. And thirdly,
for a locally isothermal disc with an $\varepsilon$ value near zero
as for the considered region, the velocity dispersions are
roughly proportional to the square root of the surface density
$( {\langle} v^2 {\rangle}^{1/2} \propto \sqrt{\sigma} )$.
If the surface density would follow the observed surface brightness,
which is nearly constant, then also the velocity dispersion would 
remain at a constant level for $R <$ 40 arcsec.
This is obviously not the case (see Fig. 2); the dispersions
neatly comply with the adopted correction to an exponential
light profile (B89).

\begin{figure}
%\vspace{11.9cm}
\psfig{figure=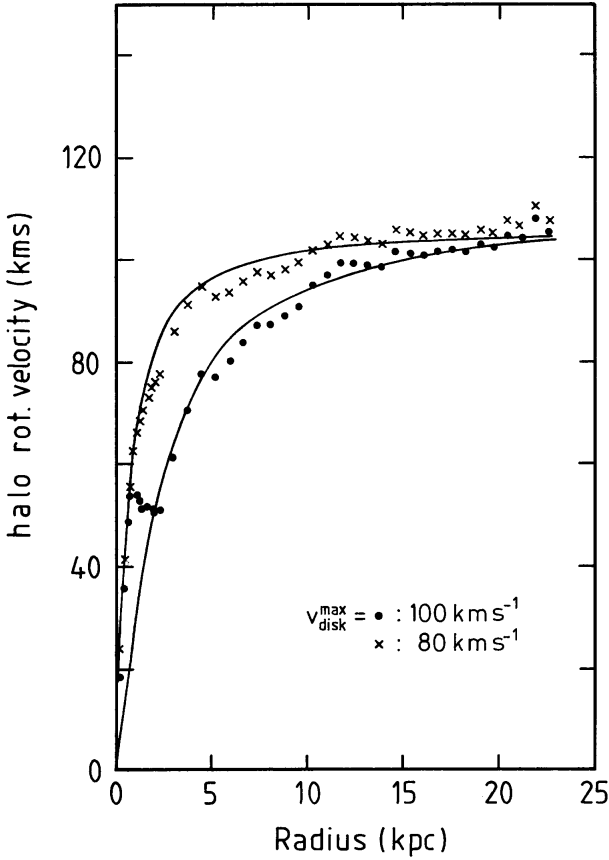,width=8.4cm}
\caption{Halo rotation curves (dots and crosses) 
derived by subtracting from the
observed rotation the rotation of a sech$^2$ disc with observed
photometry of Fig. 4 and indicated maximum disc rotational velocities.
For \vmaxds = 90 and 100 \kmss an unrealistic halo rotation results for the
inner regions. Lines are a least squares fit of Eq. (6) to the data,
with the deviant points excluded.
}
\end{figure}
It is clear that a kind of correction is necessary, although
the exact amount is difficult to establish. Certainly the adopted exponential
disc with $h = 40$ arcsec is consistent with the observed dispersions.
It is not expected that conclusions concerning global stability
are affected by this uncertainty. But there will be some leverage
when comparing predicted with observed dispersions as will be discussed
in Sect. 4.6.

At larger radii, typically for $R \ga$ 150 arcsec, there is some
additional stellar light not taken into account in the mass model
of the disc. In addition for NGC 6503 the gas mass fraction also produces 
an amount of rotation ($\la$ 29 \kms, see Begeman 1987 and Begeman
et al. 1991), which in the present calculations is included
in the halo contribution. These two approximations have a negligible
effect on the derived dark halo parameters. Also the effect on
the kinematics and morphology in the inner disc region
is close to zero.

\subsection{The halo construction}

For the adopted inwards extrapolated disc photometry
and different disc masses with different maximum disc rotations,
the resulting halo rotation curves plus fit to these are
shown in Fig.~6. The density functionality of a pseudo-isothermal
sphere is adopted for the halo:

\begin{equation}
  {\rho}_{\rmn H} = {\rho}^0_{\rmn H} \left(1 + 
  {{R^2}\over{R^2_{\rmn c}}} \right)^{-1},
\end{equation}
with

\begin{equation}
  v_{\rmn H} = \sqrt{4\pi G {\rho}_0} \; R_{\rmn c}\;
  \sqrt{1- {{R_{\rmn c}}\over{R}} {\rmn arctan}\left({{R}\over{R_{\rmn c}}}\right)
 },
\end{equation}
which generally provides a good fit to other observed
rotation curves (Carignan \& Freeman 1985). For the different \vmaxds
values the appropriate best fitting halo parameters are given
in Table~2. A fit of a dark halo model as described
by Hernquist (1990) was also attempted. Such a density
and belonging rotation curve, however,
was unable to properly fit the halo rotational data points.

\begin{table}
\caption{Disc - dark halo parameters, pure exp. disc}
\begin{tabular}{llll}
\hline
 $v^{\rmn max}_{\rmn disc}$ & Disc mass & $R_{\rmn c}^{\rmn halo}$ & 
${\rho}_0^{\rmn halo}$ \\
 (\kms) & (10$^9$ $M_{\odot}$) & (kpc) & (kg m$^{-3}$) \\
\hline
 68 & 3.49 & 0.669 & 0.367 10$^{-19}$ \\
 80 & 4.79 & 0.879 & 0.208 10$^{-19}$ \\
 90 & 6.06 & 1.26  & 0.101 10$^{-19}$ \\
 100& 7.48 & 2.16  & 0.371 10$^{-20}$ \\
\hline
\end{tabular}
\end{table}
\begin{figure}
%\vspace{8.4cm}
\psfig{figure=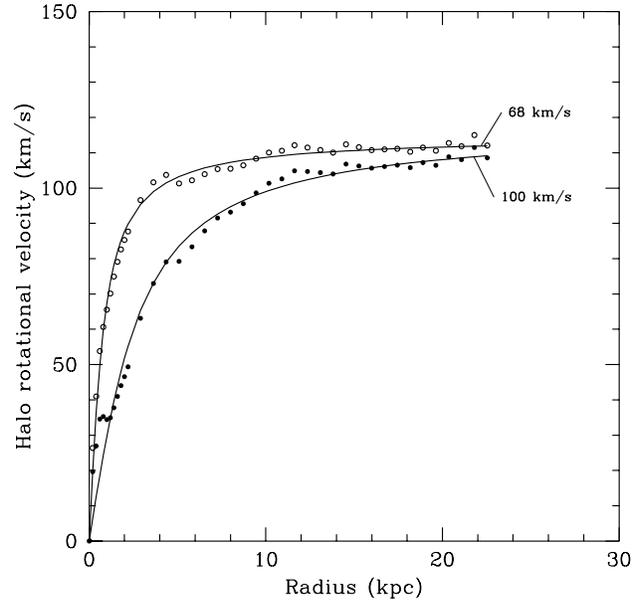,width=8.4cm}
\caption{
Halo rotation curves (dots and circles) derived by subtracting from
the observed rotation, the rotation of a sech$^2$ exponential disc with 
scalelength of 40 arcsec,
and indicated maximum disc rotational 
velocities. Lines are a least squares fit of Eq. (6) to the data,
resulting in halo parameters presented in Table 2.
}
\end{figure}
The velocity dispersions for the family of decompositions can
now be calculated. To visualize the relative importance of dark and
luminous matter the $\varepsilon$ values as they are before resettling
are shown in Fig. 7, top panel.
For radii larger
than one to two kpc the value of $\varepsilon$ is larger than zero
and consequently the velocity dispersion will somewhat increase when the
disc is embedded in the halo. This effect is stronger for the more
massive discs. 
For radii typically less than 1 kpc the value of $\varepsilon$
becomes slightly negative which might lead to lower dispersions.
In the middle panel of Fig. 7 the vertical dispersions are presented
and in the bottom panel Toomre's $Q$ value given by

\begin{equation}
  Q =  {{ \dispr \kappa }\over{3.36 G \sigma}},
\end{equation}
where the velocity dispersion in the radial direction, \dispr,
is assumed to be ${{1}\over{0.6}}$ times the dispersion in the
vertical direction and where $\kappa$ is the epicyclic frequency.
It is obvious from Fig. 7 that a more massive disc has lower
$Q$ values. Consequently such discs will succumb easier
to instabilities than the less massive discs.
\begin{figure}
%\vspace{17.3cm}
\psfig{figure=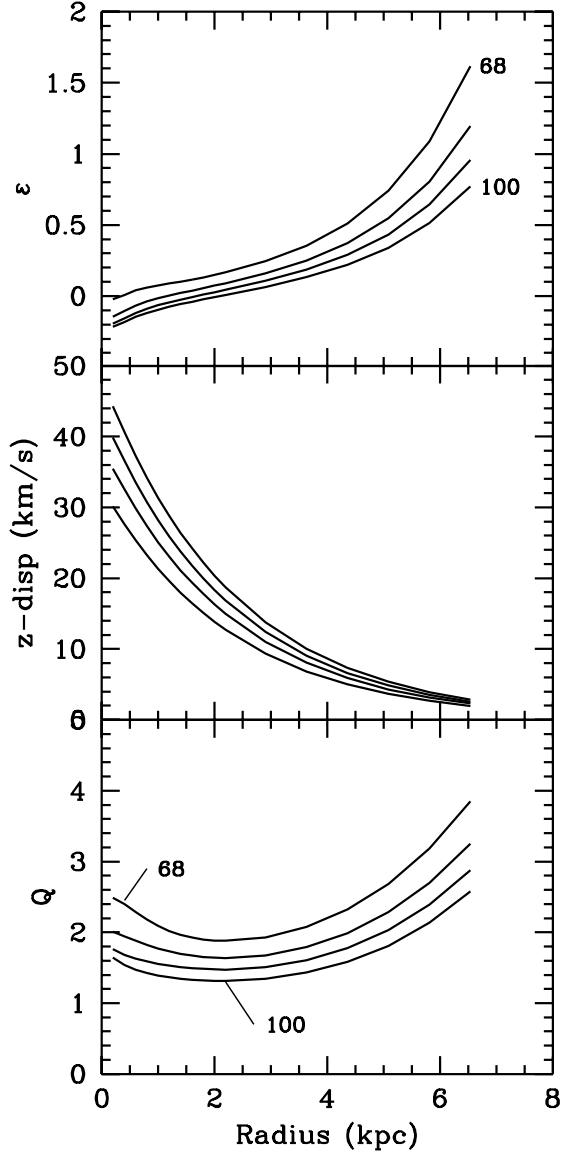,width=8.4cm}
\caption{
Predicted kinematical properties for the family of disc-halo
realizations, with \vmaxds = 68, 80, 90, and 100 \kms.
{\it Top:}
Relative contribution of disc and halo expressed in the
$\varepsilon$ value. {\it Middle:} Vertical velocity dispersion.
The upper curve, or largest dispersion, is for the most
massive disc (100 \kms). {\it Bottom:}
Toomre's $Q$ value, showing that the most massive disc is
the least stable.
}
\end{figure}

\section{Numerical simulations of NGC 6503}

\subsection{Motivation}

The initial reason for carrying out numerical simulations of NGC 6503
was to check whether a disc with
a very low central velocity dispersion can really exist. 
At the same time there are a few
other matters which can be investigated like
what is the appearance
of a stellar disc as a function of $M_{\rmn disc}/M_{\rmn halo}$ and 
can on the basis of such an appearance certain mass ratios be excluded.
And, in addition, what are the ratios of the three components of the 
velocity dispersion as a function of radius; are these ratios
consistent with those predicted by theory.

\subsection{Techniques}

In the present study the calculations
are confined to be pure stellar. Only the equilibrium and evolution
of an old stellar population can then be studied, but that is in the 
same time the population of which the dispersions have been measured.
A young population has far less absorption lines, mainly at the blue
part of the optical spectrum and has a negligible influence on the
determined dispersion in B89. In a follow up study
we have included gas and star formation into the calculations. It
appears that the effect this has on the kinematics of the
old disc population in indeed small.

The calculations were performed with a TREESPH code
(Hernquist \& Katz 1989), although for the simulations presented
here only the collisionless part of the code was used.
Dimensionless units were used, with the constant of gravity
$G = 1$. Specifically for NGC 6503 the unit of length is 1 kpc,
unit of mass is 1~10$^9$ $M_{\sun}$, unit of velocity 65.6 \kms, and
unit of time 15 Myr. For the calculation of the
gravitational forces the tolerance parameter $\theta$ was set to 0.6.
The time step in the simulations was 0.1, and energy and
angular momentum were always conserved to better than 1 per cent.
The stellar disc has been represented by 40,000 equal mass
particles with gravitational softening parameter $\epsilon$ = 20 pc.
In the present case this softening length is much smaller
than the scaleheight of the stellar disc so that there is
practically no artificial stabilization caused by large $\epsilon$ parameters
(Romeo 1994).
The halo has been represented by a fixed spherical potential.
The reason for this is, in first instance, limited computational resources.
Representing the halo by a collection of particles requires
huge numbers of particles,
distributed over a large spatial region, or a more
limited number of more massive halo particles. The latter inevitably
leads to substantial disc instabilities 
causing spurious spiral features
when such a massive
particle moves through the disc (Hernquist 1993; Lacey \& Ostriker 1985).
For our simulation where disc stability is investigated this is an 
undesirable situation. An additional reason to take a fixed halo potential
is that, until now, little is known about the phase space distribution
of halo particles. Any choice of this would be purely ad hoc, which
might lead to wrong effects and conclusions.

The disc mass was divided up in 40,000 particles. A normal galactic disc
with $\sim$ 10$^8$ to 10$^{10}$ particles is essentially collisionless.
The much smaller number
of particles in the simulation creates a situation which is not
completely collisionless, and as a consequence heating will
occur by encounters not to take place in reality.
Normally, heating is considered to take place via star-cloud
scattering (Spitzer \& Schwarzschild 1951) or by transient spiral arms
(Sellwood \& Carlberg 1984). If spiral arms occur, heating
by these is generally fast, faster than the particle scattering.
In the absence of arms we had to assume that the particle collisions
mimic, in a suitable way, the actual interactions taking
place in a real disc. Unfortunately, until simulations with much
larger numbers of particles with more realistic disc setups
can be performed, one has to do with this assumption.

\subsection{The asymmetric drift}

To facilitate the comprehension and interpretation of matters a
description of the asymmetric drift is appropriate. In a flattened
rotating system mass constituents with a certain velocity dispersion
will rotate slower than constituents with no velocity dispersion. The 
latter, testparticles in the potential, have a rotational velocity
$v_t = \sqrt{-R\; \partial\Phi/\partial R}$, which for a galaxy is nearly
equal to the rotational velocity of the gas component, which
can be observed. The Jeans equations allow the calculation of the
asymmetric drift, which for a plane parallel disc follows from

\begin{eqnarray}
  v_t^2 - v_{\ast}^2 &=&{\langle}v^2_R{\rangle} \left[ {{-R}\over{\rho}}
  {{\partial}\over{\partial R}} \rho (R) - {{R}\over{{\langle}v^2_R{\rangle}}}
  {{\partial}\over{\partial R}} {\langle}v^2_R{\rangle} \right] -
  \nonumber \\
  & &{\langle}v^2_R{\rangle} \left[ \left( 1 - 
  {{{\langle}v^2_{\Theta}{\rangle}}\over{{\langle}v^2_R{\rangle}}} 
  \right) \right],
\end{eqnarray}
(Oort 1965; Binney \& Tremaine 1987 page 199). The drift,
$\Delta v = v_t - v_{\ast}$ is then for the isolated exponential disc used
as input to the simulations

\begin{equation}
  2v_t \Delta v \approx v_t^2 - v^2_{\ast} =
  {\langle}v^2_R{\rangle}\; \left[ {{2R}\over{h}} + {{1}\over{2}}\left(
  {{R}\over{v_{\ast}}} {{\partial v_{\ast} }\over{\partial R}} 
  - 1 \right) \right],
\end{equation}
with $v_{\ast}$ denoting the rotation of the stars. 
This poses a slight problem. 
If $v_{\ast}$  could be replaced by $v_t$, the
testparticle rotation (which is the same for all
disc/halo mass decompositions) then $\Delta v \propto {\langle}v^2_R{\rangle}$.
For the different disc masses the asymmetric drift would scale with
the velocity dispersion in that disc; and this scheme was initially
adopted to give the input stellar rotation. But 
in calculating the asymmetric drift terms like
$\partial v_{\ast} / \partial R$ are involved which may, in the central region,
differ considerably from $\partial v_t / \partial R$ and will change
for the different disc masses.
The asymmetric drift should be calculated from the stellar rotation which is
not known in advance and therefore, starting out with the rotation
of the ionized gas, always an iteration is involved towards the true
stellar rotation.

\subsection{Setup and settling}

The initial density structure of the galaxy was
made equal to one of the disc/halo realizations
as described in the previous section.
Initial dispersions were chosen to be
representative for an isolated locally
isothermal exponential disc:

\begin{equation}
  \dispz = \sqrt{\pi G \sigma(R) z_0} \propto {\rmn e}^{-R/2h}
\end{equation}
\begin{equation}
  \dispz/\dispr = 0.6
\end{equation}
\begin{equation}
  \dispt/\dispr = \sqrt{B/(B-A)}.
\end{equation}
Mind that such a prescription is a reasonable approximation of,
but not completely valid for a real galactic disc.

In first instance the observed gas rotation was used as input
rotation for the stars. It was expected that the disc would settle
gently towards the kinematics appropriate for the galaxy. However,
the result was an outflow of sizeable amounts of stars from the
central 1 kpc region altering the surface density and dispersions in an
undesirable way.
In second instance the input stellar rotation was taken to be
that of the gas diminished with the asymmetric drift calculated
from the gas rotation curve (Eq. 9 with $v_{\ast}$
replaced with $v_t$).
Still for all disc masses central outflow occurred.
For \vmaxds of 68 and 80 \kmss one iteration, as explained in the 
previous subsection, was performed
on the input stellar rotation to make the outflow as small
as possible. However, for the larger disc masses this was more difficult
or impossible to accomplish because in the settling process there
are always small changes in the density, rotation, and dispersions.
Especially for $R <$ 1 kpc the asymmetric drift is extremely sensitive
to such changes, making it difficult
to successfully do the iteration to a stable stellar kinematics.
Hence certainly for the more massive discs
some initial outflow was taken for granted.

For the lightest disc the drift is small, at most a few \kms.
For \vmaxds = 80 \kmss it reaches values between 8 to 15 \kmss for 
$R <$ 1 kpc. But the asymmetric drift is observed! In B89 the
rotation of the stars and H$\beta$ gas are compared and the 
observed difference is less than 7 \kmss over this radial extent.
Even if one were to disbelieve the measured velocity dispersions
indicative of a less  massive disc, this is an independent prove
that \vmaxds $\la$ 70 \kms.

For the settled situation the testparticle velocity was calculated from the
density distribution. A stellar rotation was monitored directly and 
hence an asymmetric drift was always given by the simulations. It was
satisfactory to see that the value of the drift was always to within
1 to 2 \kmss of the theoretical expectation from Eq. (8).

\begin{figure*}
%\vspace*{18cm}
\psfig{figure=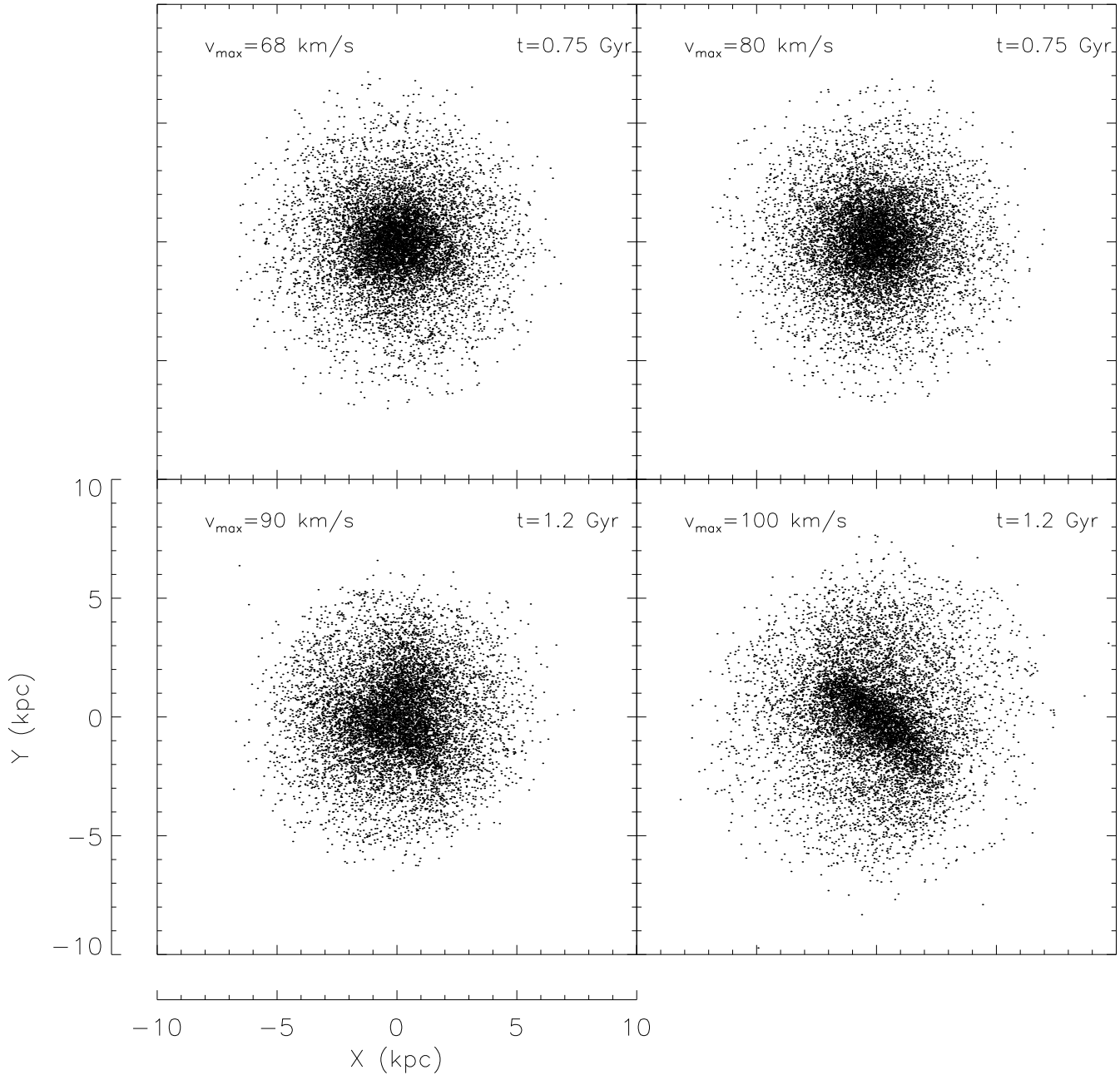,width=17.8cm}
\caption{
Face-on appearance of the simulated stellar disc representative for
that of NGC 6503. Of the 40,000 particles only one in four is plotted.
The two situations with most massive disc exhibit large scale instabilities,
not compatible with the observed appearance (Fig. 1) of NGC 6503.
}
\end{figure*}

\subsection{Results}

The simulations were continued for at least 7.5 10$^8$ years, and in all
cases (\vmaxds = 68, 80, 90, and 100 \kms) the disc was settled to a steady
situation in about half this time. For the two lighter discs the face-on
appearance during the whole simulation is essentially featureless, with
only a hint of spiral arms for the \vmaxds = 80 \kmss situation. In the
case of \vmaxds = 90 \kmss there is initially some resettlement of stars caused
by the small central outflow. Then, after approximately 4.5 10$^8$ yrs the
inner half of the galactic disc develops a triangular structure, now and
then disappearing and appearing again after continued simulation. 
For \vmaxds = 100 \kmss the disc
is highly unstable evolving to the same triangular structure at
t = 7.5 10 $^8$ yrs, which then quickly settles to a stable bar which 
dominates most of the stellar disc region. Continued simulation
shows that it persists for at least 2.5~10$^9$ years, finally
decreasing in strength because of the strong heating of the disc
in the outer regions. This heating also results in an increasing
thickness of the disc for larger radii contrary to the situation
of the lighter discs. During the simulation
the stars slowly heat up caused by
particle scattering for the \vmaxds = 68 and 80 \kmss cases.
Likely this heating for the more massive discs is additionally generated
by the observed transient features and/or the bar, but this has
not been investigated any further.
For the four different disc masses the face-on density distribution
is presented in Fig. 8 at or beyond a time of 7.5 10$^8$ years.
After comparison
with the observed highly regular structure of NGC 6503 we conclude that 
a disc contributing more than 90 \kmss to the maximum observed
rotation of 120 \kmss is ruled out. A situation near that of 
``maximum disc'' (van Albada \& Sancisi 1986) is not possible
just by comparing simulated and observed density distributions.
Rhee (1996) gives a maximum disc rotation of 108 \kmss in
the case of this maximum disc limit. Then, in order to avoid a bar,
an actual disc of NGC 6503 has at most a mass of 70 per cent of the mass of that 
limiting case.

As noted above, spiral arms are essentially absent. This is to be
expected because gas is not included in the experiments and galaxies
without gas never show spiral structure.
Only by cooling of the gas, spiral arms can appear in that component
and by additional star formation the young ``cool'' stars will
generate an accompanying spiral disturbance in the stellar population
(Sellwood \& Carlberg 1984; Carlberg \& Sellwood 1985).
It is not to be expected that the absence of gas will considerably
influence the results concerning stability against global
bar formation since such a disturbance is one of the whole 
massive stellar disc.

\begin{figure*}
%\vspace*{21cm}
\psfig{figure=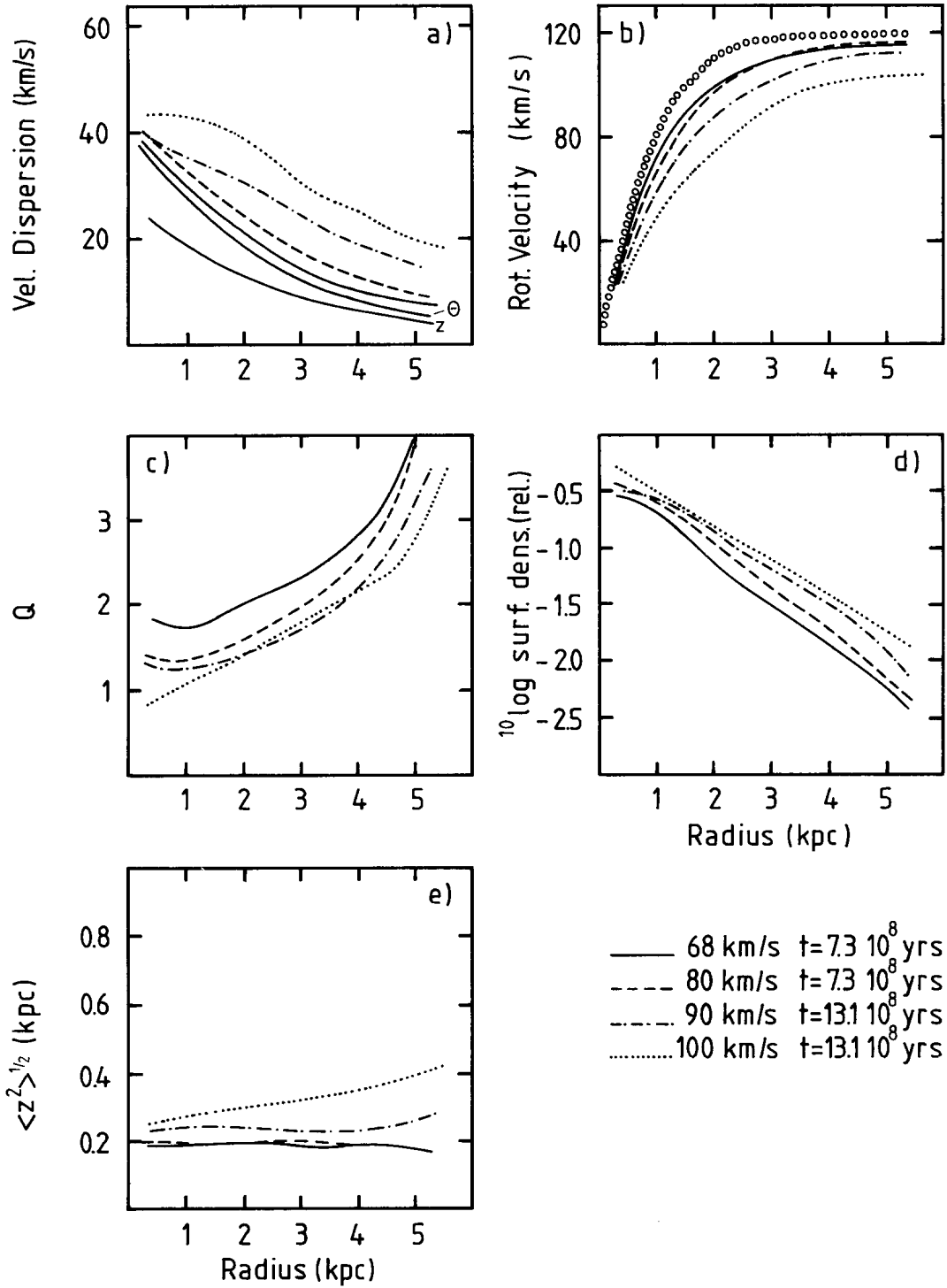,width=15cm}
\caption{
A number of galactic properties of the simulated galaxy for different
disc contributions. {\bf a} Radial velocity dispersion ($+$ vertical and
tangential for \vmaxds = 68 \kms). {\bf b} Stellar rotational velocity. 
Also indicated by the circles is the zero dispersion rotation for the
\vmaxds = 68 \kmss situation at t = 7.3~10$^8$ yrs to show the asymmetric
drift. {\bf c} Toomre's $Q$ value. {\bf d} Surface density in relative
units. {\bf e} Rms z-height.
}
\end{figure*}

In Fig. 9 the resulting values of the velocity dispersion (9a),
stellar rotation (9b), $Q$ value (9c), surface density (9d),
and rms z-height (9e) are presented.
Only for the two lighter discs the input rotation was adjusted to 
diminish the central outflow as much as possible.
In Fig. 9d the final radial density profile indicates how much
mass was lost from the centre. Because only a small area
is concerned the lost mass fraction is always negligible compared to
the total mass of the stellar disc.
It can be seen in Fig. 9b that the asymmetric drift increases rapidly
for the larger disc masses, especially at intermediate and large radii.
As noted in the previous subsection the drift which is equal to the difference
between the stellar rotation curves and the test particle rotation curve 
in Fig. 9b is equal to within 1 to 2 \kmss to that
predicted by Eq. (8). The asymmetric drift can, in principle, be observed 
in external galaxies by comparing the rotation velocities of the stars
to that of the ionized gas. Unfortunately because of the low
surface brightness of stellar discs such observations are confined
to radii $\la 1{{1}\over{2}}$ to 2 scalelengths. 
But maybe with the advent of large
telescopes, in the future such measurements might be feasible; the
drift in the outer regions is indeed very sensitive to the local
disc mass density values.

The situation was started out with a z-velocity dispersion expected 
for that of an isolated stellar disc. During the simulations this
dispersion, as well as the $R$ and $\Theta$ dispersions, gradually increase
partly caused by particle scattering.
At the larger radii there 
will be an additional small increase of
the dispersion because the disc is settling to an equilibrium
situation, as descussed in Sect. 2. One would also
expect the disc to become thinner but this in not observed.
As can be seen in Fig. 9e the rms scaleheight remains practically
constant as a function of radius.
This might be explained by a more violent particle scattering
in the outer parts, therewith heating the disc more than in the
inner regions and increasing the scaleheight. At present it is still
not understood why for practically all spiral galaxies 
the observed scaleheight is so constant. These
simulations suggest that this constancy might come about by
extra scattering of stars in the outer regions by some mechanism.

The small additional heating in the outer parts also generates 
a somewhat larger $Q$ value at large $R$ (Fig. 9c) compared to what
one expects for an isolated exponential disc having $\dispz / \dispr = 0.6$
as given in Fig. 7. Nevertheless, on the whole, the simulated and
predicted $Q$ values agree very well, even though
at the end of the simulations the disc is not exactly exponential
any more and the ratio of vertical to radial dispersion need
not to have remained at the same constant level.
For the
non axisymmetric \vmaxds = 100 \kmss case a $Q$ stability value does
not really apply. Still also for that case all parameters have been calculated
rigorously, azimuthally averaged.
For \vmaxd = 68 and 90 \kmss the simulations have been repeated
with 200,000 particles instead of the default number of 40,000.
The appearance of the disc does not change when increasing
the number of particles. Heating of the disc and the
increase of the disc thickness is less fast, as one might expect.
But the radial functionality of the dispersions and $Q$
value do not change in a significant way.

As input for the simulations a vertical to radial dispersion ratio
of 0.6 was taken based on observations of old disc stars in the
solar neighbourhood and on work by Villumsen (1985) and B93.
At the end of the simulations there has been little evolution
from this value of 0.6 for all four disc mass situations. Typically within
a radius of two scalelengths the $z/R$ dispersion ratio becomes a bit
larger reaching $\sim$ 0.65; while in the most outer regions of the disc values
decrease to around 0.5. The mechanism of stellar heating in a galactic
disc is still poorly understood. Heating by molecular clouds can
at least do part of the job
(Spitzer \& Schwarzschild 1953; Icke 1982; Lacey 1984; 
Binney \& Lacey 1988; Villumsen 1985) but not at the right pace.
On the other hand, heating by transient spiral arms is more efficient
(Sellwood \& Carlberg 1984) but by this mechanism only radial and tangential
dispersion heating can be explained. 
The dispersions following from the present simulations
are generated in a situation where particle
scattering is important. This is  not the same
as for the actual galaxy
and therefore a detailed comparison between simulated and
observed dispersions cannot be made.
Still the present simulations at least indicate that
a ratio of 0.6 for the $z$ to $R$ dispersion ratio
is close to the stable equilibrium situation
for a disc embedded in a dark halo.
Also the simulated tangential to radial dispersion ratio is compared
with what one might expect following the Jeans equations 
(Binney \& Tremaine 1987, Eq. 4.51).
There is some discrepancy for $R$ $\ga 1{{1}\over{2}}$ kpc in the sense that
the simulated value is larger by $\sim$ 0.1. 
Viewed in the light of the discussion above,
such differences can be expected.

\subsection{Comparison with the observed dispersions}

For NGC 6503 with an inclination of 74\degr the observable
dispersion is nearly equal to the velocity dispersion in the
tangential direction (\dispt). This is caused by the fact
that in the integrated line-of-sight dispersion a somewhat smaller
z-dispersion entering by a not edge-on situation is
approximately compensated for by a somewhat larger R-dispersion
entering because of the thickness of the stellar layer.
Hence the tangential dispersion of the
simulation can be compared with the observed dispersions
directly as shown in Fig. 10. From this comparison
it is obvious that the more massive discs are excluded;
the observed dispersions are simply too small.
This conclusion was already reached for a general galactic
disc in B93.

\begin{figure}
%\vspace{8.4cm}
\psfig{figure=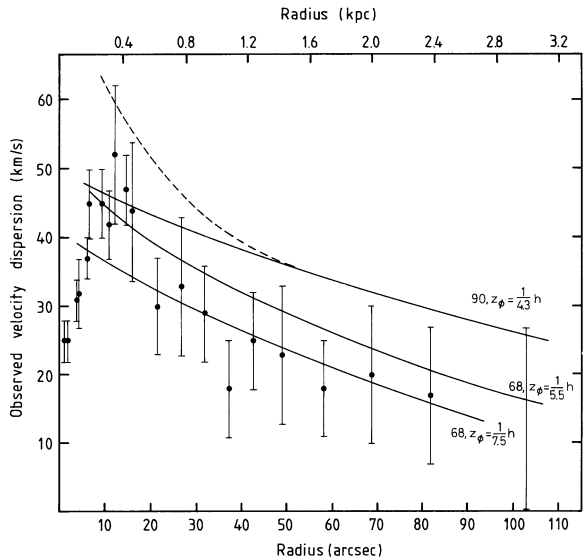,width=8.4cm}
\caption{
Simulated tangential dispersion compared with the observed stellar velocity
dispersions of NGC 6503. Lines are for \vmaxds = 68 and 90 \kmss and
indicated $h/z_0$ value. For the 90 \kmss case some mass
was lost from the centre resulting in a lower local dispersion.
A correction towards the initial exponential disc is indicated by the 
dashed line. The comparison clearly
shows that the more massive discs are excluded. Unfortunately
the central dispersion drop cannot be reproduced by the simulations.
}
\end{figure}

Some care has to be taken for situations where mass was lost from the
central regions; the dispersion is lower there because the surface density
is smaller. A simple correction has been made for this in Fig. 10,
by pretending the surface density remains exponential
into the centre and scaling the dispersion with the square root of the ratio
of corrected to uncorrected density. This is indicated by the dashed line.
For the \vmaxds = 68 \kmss case the simulated dispersion
values are a bit larger than what is
observed.
Smaller dispersions can be obtained by decreasing the scaleheight
as is shown in Fig. 10 by a simulation with \vmaxds = 68 \kmss
and $h/z_0$ = 7.5. For that case an excellent fit is found for
$R >$ 0.6 kpc, but data points between 0.3 and 0.6 kpc
are too large by $\sim$ 10 \kms. When doing such comparisons it
should be taken into account that the surface density
for the real galaxy is somewhat uncertain because of the poorly
defined absorption correction. 
Additionally, stellar heating mechanisms differing locally
can cause a non perfect match between observations and simulations.

However, the drop in dispersions for $R <$ 0.2 kpc as observed
is not reproduced
in the simulations. It has been investigated whether an initial
situation with a very cool disc, might, after some evolution
and heating, leave a cool centre behind. But that did not appear; the
centre heats up and the thickness increases 
at the same rate as the rest of the disc.
At least all the possibilities we investigated
did never lead to a cool centre,
which does not rule out completely that
it cannot be achieved.
Still one can conclude that it is very difficult to
construct a lasting cool centre for an exponential
sech$^2$ disc.

\subsection{Concluding}

To summarize this section, it can be concluded that
a massive disc, approaching the maximum disc limit
is unstable to bar formation. In addition, for such a situation the galactic
disc has larger dispersions and larger asymmetric drifts
than is observed.
It is, however, hard to explain the observed dispersion drop 
at the centre of NGC 6503 by a kinematic construction in a pure
stellar disc.

\section[]{The explanation of the central\\* dispersion drop}

\subsection{Introduction}

The analyses in the previous sections made it clear that
the dispersion drop is actually very centrally confined; 
to a radial extent less than $\sim z_0 \sim$ {6}\farcs{7}.
This region coincides with a central peak in the luminosity
profile. Represented in magnitudes this
luminosity peak is not impressive (Fig. 4), but on a true
linear scale it is dominant.
Hence in the centre there is a component much brighter
and much smaller than the surrounding disc. It might be that
the disc density and density gradient over the peak region are
small or even negligible. If that is the case one might consider
the central component as a separate isolated entity. Therefore
the nucleus will be assumed to be an isolated isothermal sphere;
any rotation is not taken into account.
Later on we will investigate whether this assumption is valid.

\subsection{Nucleus-disc light decomposition}

The same absorption correction as before has been applied to the
central regions. An exponential disc with a scalelength of 40 arcsec has
been subtracted from the total light and the remainder is considered
to be the light contribution of the nucleus. To the light profile
of the nucleus a seeing correction has been applied for a FWHM seeing
of 2 arcsec and the resulting profile is presented in Fig. 11.
The approximate total light of the nucleus was found by fitting an 
exponential to the light profile and integrating this fit. Results
a total light of 7.5 10$^7$ $L^{\rmn R}_{\sun}$ or an absolute magnitude
in the R-band of -15.21. 
For the observed B-R colour of 1.24 the absolute magnitude
in B is -13.97.

\begin{figure}
%\vspace{8.6cm}
\psfig{figure=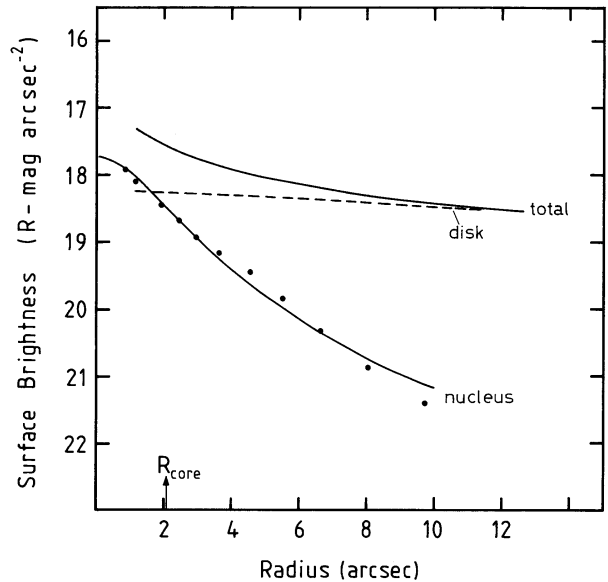,width=8.4cm}
\caption{
Light decomposition into the inwards extrapolated exponential disc and
nucleus (dots). The line is a fit of a modified Hubble profile to the
nuclear data and resulting core radius is indicated.
}
\end{figure}

\subsection{The core fitting method}

King's core fitting method will be applied to the nucleus
(King 1966, 1981; Richstone \& Tremaine 1986).
Usually from the observed central dispersion, surface brightness,
and core radius a M/L ratio is derived. In the
present case the method is inverted such that it is investigated whether
the nucleus with the same M/L as the disc can explain the observed
central velocity dispersion.

Up to $R \sim 3R_{\rmn core}$ a modified Hubble profile

\begin{equation}
  \Sigma(R) = {{ {\Sigma}_0}\over{ 1 + 
  \left( {{R}\over{R_{\rmn core}}} \right)^2 }},
\end{equation}
is representative for the projected light distribution
$\Sigma(R)$ of an isothermal sphere. Eq. (13) has been fitted to the 
nuclear light profile in Fig. 11, resulting in a central surface
brightness ${\Sigma}_0$ of 17.74 $\pm$ 0.1 R-mag per square arcsec
(= 1796.4 $L_{\odot}$ pc$^{-2}$) and a core radius of 60.91 $\pm$ 4 pc.
For the deprojection of an isothermal sphere then
follows a central luminosity density $j_0$

\begin{equation}
  j_0 = {{0.495 {\Sigma}_0}\over{R_{\rmn core} }},
\end{equation}
of 14.60 $\pm$ 2.2 $L_{\sun}$ pc$^{-3}$. To calculate the
velocity dispersion the central luminosity density will be converted
to the central mass density ${\rho}^{\rmn c}_0$ by assuming that the
nucleus has the same M/L ratio as the disc.
The mass values of the disc are derived from the observed disc
velocity dispersions, which imply \vmaxds = 70 $\pm \sim$ 10 \kms.
For an extrapolated exponential disc inwards to $R = 0$ this gives
${\sigma}^{\rmn disc}_{R = 0}$ = 460 $M_{\sun}$pc$^{-2}$. The disc
light is given by the photometry and nucleus-disc light decomposition
(Fig. 11). For the exponential disc ${\mu}^{\rmn R}_{R=0}$ = 18.2 mag
per square arcsecond = 1180.3 $L_{\sun}{\rmn pc}^{-2}$. For a simple
plane parallel correction to face-on this results in a face-on
surface brightness of 325 $L_{\sun}^{\rmn R}{\rmn pc}^{-2}$.
Then $(M/L)^{\rmn R}_{\rmn disc} = {\sigma}^{\rmn disc}_{R=0}/
{\mu}^{\rmn face-on}_{R=0} = 1.41\; M_{\sun}/L_{\sun}^{\rmn R}$.
The considerable error in the observed disc velocity dispersions
and the rather uncertain inclined to face-on brightness correction
result in an error of $(M/L)_{\rmn disc}$ of approximately 
a factor of two.

For $(M/L)_{\rmn disc} = (M/L)_{\rmn nucleus}$ the central mass density of
the nucleus $({\rho}^{\rmn c}_0)$ is 20.6 $M_{\sun}{\rmn pc}^{-2}$
($\pm$ factor of 2). The velocity dispersion of the nucleus
$({\sigma}_{\rmn nuc})$ is given by

\begin{equation}
  {\sigma}_{\rmn nuc} = \sqrt{
  {{ 4 \pi G R^2_{\rmn core} {\rho}_0^{\rmn c} }\over{9}} },
\end{equation}
and amounts to 21.5 \kms $\pm$ 35 per cent.

A rather ad-hoc dust absorption correction has been
applied to the inner regions and one might question the influence
of that.
Therefore the calculation of the dispersion of the nucleus has been
repeated for the photometric profile as actually observed. The
result is nearly identical, showing that the effect of the absorption
correction on the resulting dispersion value is negligible.
This can be comprehended as follows: for a less severe
correction as has been applied, the nucleus will be
less bright, but the
M/L ratio of the disc is increased by nearly the same factor. Hence
the central mass density of the nucleus remains nearly equal giving
the same velocity dispersion (Eq.~15). It should be noted that
the calculated dispersion value does not depend
on the adopted distance to the galaxy.

\subsection{The influence of disc and halo}

For the nucleus until $R \sim 3R_{\rmn core}$ the fitted modified Hubble
profile of Eq. (13) can be deprojected to give the space density
${\rho}_{\rmn nuc}$ as a function of internal radius $r$

\begin{equation}
  {\rho}_{\rmn nuc}(r) = {{ {\rho}_0^{\rmn c} }\over
  { \left[ 1 + \left( {{r}\over{R_{\rmn core}}}\right)^2 \right]^{3/2} }}.
\end{equation}
The space density of the disc at the centre of the
galaxy is given by Eq. (1) with ${\rho}_{\rmn disc}(0,0) = 1.19\;
M_{\sun}{\rmn pc}^{-3}$ and $z_0$ = 194 pc. For the dark halo see 
Eq. (5) with in the present case, for \vmaxds = 68 \kms, 
${\rho}_0^{\rmn halo} = 0.54\; M_{\sun}{\rmn pc}^{-3}$ and
$R^{\rmn halo}_{\rmn core}$ = 669 pc. At $R = 0$ the densities of the
three components are compared, as a function of z-height in Fig. 12.
From this figure it is obvious that the nucleus is dominant
in the centre. It is completely embedded in the disc, the disc plus
halo density is nearly constant in the region of the nucleus.
Hence the assumptions of Sect. 5.1. are valid and an isolated 
isothermal sphere provides a good description of the actual situation.

\begin{figure}
%\vspace{10.2cm}
\psfig{figure=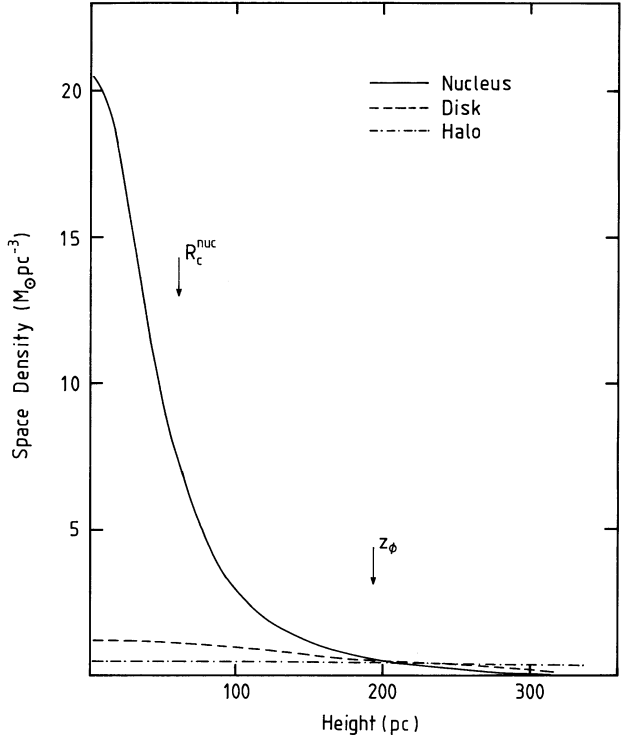,width=8.4cm}
\caption{
At $R = 0$, densities of nucleus, disc and halo as a function of
height above the plane. Assumed are equal mass-to-light ratios of
nucleus and disc and \vmaxds = 68 \kms. Note that the nucleus
is dominant and completely embedded in the disc.
}
\end{figure}

The dispersion of the nucleus has been determined at 21.5 \kms, but the
actually observed dispersion along the major axis (Fig. 2) is a
combination of nucleus and disc kinematics. To see if the observations
can actually be reproduced, a kinematical model of NGC 6503
has been constructed in the same way as in Bottema (1995).
The galaxy consists of a disc and a nucleus. The disc is exponential
($h$ = 40 arcsec) with vertical distribution of Eq. (1) and rotating
with velocities given by Fig. 3. The velocity dispersions were according
to equations (10) to (12) with a central radial dispersion of 55 \kms.
The bulge was spherical with density distribution given by Eq. (16),
relative density with respect to the disc as given above,
cylindrically rotating with the same rotation curve as the disc, and
having an isotropic dispersion of 21.5 \kms.
Line profiles were calculated numerically for the combined
nucleus-disc situation with an inclination of 74\degr, and 
gaussians were fitted to these profiles. The dispersion determined
in this way can be compared with the dispersions following from the
observations, which has been done in Fig. 2. It appears that
the data given by the nucleus-disc model are in good agreement
with the observations, not only in magnitude but also at the right
radial positions. This proves that both, the light decomposition and
dispersion calculations are consistent with the actual situation.
For a different absorption correction the dispersion along the
major axis remains equal. This is because the relative light
contributions of nucleus and disc hardly change, and also, as noted
above, the nuclear dispersion remains the same.

\subsection{Conclusion}

There is an excellent agreement between the calculated dispersion and the
actual observed value of the nucleus. The explanation by considering
the nucleus as a separate galactic component appears to be right.
The core fitting procedure can also be turned around now and used
in its original form. Then, if the observed central dispersion is that
of the nucleus it can be concluded, taking into account the errors,
that the mass-to-light ratios of nucleus and disc are equal to
within a factor of two.

\subsection{Discussion}

Kormendy \& Illingworth (1983) compare
the observed dispersions and brightnesses of several bulges. Their graph
is reproduced here in Fig. 13 with the value of the nucleus of NGC 6503
included. On the basis of the fitted relation of Kormendy \& Illingworth, 
the straight line in Fig. 13,
one expects a dispersion for NGC 6503 of 90 - 100 \kms.
Instead $\sim$ 25 \kmss is observed. However the fit was performed
over a narrow range of bulge brightnesses and should be considered 
uncertain. To fill up the range between NGC 6503 and the other bulges
one needs observations of (small) dispersions of small bulges. 
One candidate was found by us: the bulge of the face-on Sc galaxy NGC 3938.
A velocity dispersion of 40 \kmss was measured by Bottema (1988). The
presented light intensity along the spectrograph slit for the central
region and photographic photometry by van der Kruit \& Shostak (1982)
for the outer regions enabled a bulge to disc light decomposition.
For the resulting $L_{\rmn bulge}/L_{\rmn total}$ = 0.07 and absolute
magnitude in B of -19.66 (Sandage \& Tammann 1981) the absolute magnitude
of the bulge of NGC 3938 amounts to -16.77. This data point has been 
included in Fig. 13 and adds to the evidence that the fit
by Kormendy \& Illingworth cannot be extrapolated to small bulges.

\begin{figure}
%\vspace{6.2cm}
\psfig{figure=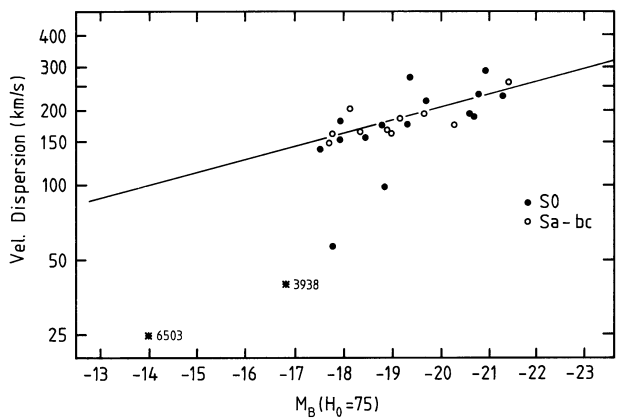,width=8.4cm}
\caption{
Bulge velocity dispersions versus bulge absolute luminosity (in B), 
reproduced from Kormendy \& Illingworth (1983). Added are the two small
bulges of NGC 6503 and NGC 3938. The line is a fit by 
Kormendy \& Illingworth to their data points, which is not in accordance
with the values of NGC 6503 and 3938.
}
\end{figure}

There is some ambiguity concerning the nomenclature of the central ``thing''.
We have used the term nucleus throughout, but it could also be called a 
small bulge. Summarizing its observed and/or inferred parameters:
a velocity dispersion, core radius, and mass of respectively 21.5 \kms,
64 pc, and 1.06~10$^8$ $M_{\sun}$.

\section*{Acknowledgments}

We thank L. Hernquist for kindly providing the TREESPH code and R. Sanders
and C. Lacey for helpful discussions and criticism. 
The investigations were supported (in part) by the
Netherlands Foundation for Research in Astronomy (NFRA) with financial aid
from the Netherlands organization for scientific research (NWO). R.B. thanks
the Kapteyn Institute for hospitality and support.

\label{lastpage}
\end{document}